\documentclass[12pt]{article}
\usepackage[cp866]{inputenc}
\usepackage[english,russian]{babel}

\renewcommand{\d}{{\rm d}}

\begin{document}

\title{First order representation of the Faddeev formulation of gravity}
\author{V.M. Khatsymovsky \\
 {\em Budker Institute of Nuclear Physics} \\ {\em
 Novosibirsk,
 630090,
 Russia}
\\ {\em E-mail address: khatsym@inp.nsk.su}}
\date{}
\maketitle
\begin{abstract}
We study Faddeev formulation of gravity, in which the metric is composed of vector fields or the tetrad of the ten-dimensional fields, $f^A_\lambda$, where $\lambda = 1, 2, 3, 4$ and $A = 1, \dots, 10$ is vector index w. r. t. the Euclidean (or Minkowsky) ten-dimensional spacetime. We propose representation of the type of the Cartan-Weyl one. It is based on extending the set of variables by introducing the infinitesimal SO(10) connection. Excluding this connection via equations of motion we re\-pro\-du\-ce the original Faddeev action. A peculiar feature of this representation is occurrence of the local SO(10) symmetry violating condition so that SO(10) symmetry is only global one in full correspondence with that the original Faddeev formulation just possesses SO(10) symmetry w. r. t. the global SO(10) rotation of the Euclidean ten-dimensional spacetime. We also consider analog of the Barbero-Immirzi parameter which can be naturally introduced in the considered re\-pre\-sen\-ta\-ti\-on.
\end{abstract}

keywords: Einstein theory of gravity; connection; Cartan-Weyl formulation

PACS numbers: 04.20.-q

MSC classes: 83C99; 53C05







\section{Introduction}

For simplicity, we consider the case of the Euclidean signature. Faddeev has con\-si\-de\-red \cite{Fad1}-\cite{Fad4} a set of ten 4-vector (or four 10-vector) fields $f^\lambda_A$. Here Greek indices $\lambda, \mu$, \dots = 1, 2, 3, 4, Latin capitals $A, B$, \dots = 1, \dots, 10. The metric tensor is
\begin{equation}                                                            
g_{\lambda\mu} = f^A_\lambda f_{\mu A}.
\end{equation}

\noindent The Latin capitals label coordinates of an Euclidean ten-dimensional space. The Greek indices label the indices of the four world coordinates and can be raised/lowered in the usual way with the help of the metric tensor. The value is introduced,
\begin{equation}                                                            
\Omega_{\lambda, \mu\nu} = f^A_\lambda f_{\mu A, \nu}, ~~~ \Omega^\lambda_{\mu\nu} = g^{\lambda\rho} \Omega_{\rho, \mu\nu},
\end{equation}

\noindent which has the same transformation properties at diffeomorphisms as connection does. The comma in indices usually means derivative, $f_{\mu A, \nu} \equiv \partial_\nu f_{\mu A}$, with exception of special definitions like $\Omega_{\lambda, \mu\nu}$ and torsion and curvature just below. The torsion is
\begin{equation}                                                            
T_{\lambda, [\mu\nu]} = \Omega_{\lambda, \mu\nu} - \Omega_{\lambda, \nu\mu}, ~~~ T^\lambda_{[\mu\nu]} = g^{\lambda\rho} T_{\rho, [\mu\nu]}.
\end{equation}

\noindent The curvature tensor is
\begin{equation}                                                            
S^\lambda_{\mu, \nu\rho} = \Omega^\lambda_{\mu\rho, \nu} - \Omega^\lambda_{\mu\nu, \rho} + \Omega^\lambda_{\sigma\nu} \Omega^\sigma_{\mu\rho} - \Omega^\lambda_{\sigma\rho} \Omega^\sigma_{\mu\nu}
\end{equation}

\noindent or
\begin{eqnarray}                                                            
& & S^\lambda_{\mu, \nu\rho} = \Pi^{AB} (f^\lambda_{A, \nu} f_{\mu B, \rho} - f^\lambda_{A, \rho} f_{\mu B, \nu}), \nonumber \\
& & \label{S lambda mu nu rho} S_{\lambda\mu, \nu\rho} = \Pi_{AB} (f^A_{\lambda, \nu} f^B_{\mu, \rho} - f^A_{\lambda, \rho} f^B_{\mu, \nu})
\end{eqnarray}

\noindent or
\begin{equation}                                                            
S_{\lambda\mu, \nu\rho} = b^A_{\lambda \nu} b_{A \mu \rho} - b^A_{\lambda \rho} b_{A \mu \nu}
\end{equation}

\noindent where
\begin{equation}                                                            
\Pi_{AB} = \delta_{AB} - f^\lambda_A f_{\lambda B}
\end{equation}

\noindent is projector onto the directions in the ten-dimensional space orthogonal to subspace spanned by the four 10-vectors at the given point ("vertical" directions) and
\begin{equation}                                                            
b^A_{\lambda \mu} = \Pi^{AB} f_{\lambda B, \mu}.
\end{equation}

\noindent The action is
\begin{equation}\label{Fad action}                                          
\int {\cal L} \d^4 x = \int g^{\lambda \nu} g^{\mu \rho} S_{\lambda \mu, \nu \rho} \sqrt {g} \d^4 x = \int \Pi^{AB} (f^\lambda_{A, \lambda} f^\mu_{B, \mu} - f^\lambda_{A, \mu} f^\mu_{B, \lambda}) \sqrt {g} \d^4 x.
\end{equation}

\noindent The variation of action is represented as
\begin{equation}                                                           
\delta \int {\cal L} \d^4 x = 2 \int (H_{\lambda \mu} f^\mu_A + \Pi_{AB} V^B_\lambda ) \delta f^{\lambda A} \d^4 x.
\end{equation}

\noindent Here $\Pi_{AB} V^B_\lambda$ are vertical and $H_{\lambda \mu} f^\mu_A$ are horizontal components of the variation w. r. t. $f^{\lambda A}$. The calculation gives
\begin{equation}\label{V lambda A}                                         
V_{\lambda A} = b^\mu{}_{\mu A} T^\nu_{[\lambda \nu]} + b^\mu{}_{\lambda A} T^\nu_{[\nu \mu]} + b^\mu{}_{\nu A} T^\nu_{[\mu \lambda]}.
\end{equation}

\noindent The equations of motion $V_{\lambda A} = 0$ give $T^\lambda_{[\mu\nu]} = 0$ almost everywhere in the infinite-dimensional configuration superspace whose points are functions $f^\lambda_A (x), x \in R^4$. Then the connection $\Omega_{\lambda, \mu\nu}$ is Riemannian one $\Gamma_{\lambda, \mu\nu}$, the curvature tensor $S^\lambda_{\mu, \nu\rho}$ is Riemannian one $R^\lambda_{\mu, \nu\rho}$ and the ten independent components of the horizontal part of the equations of motion coincide with the Hilbert-Einstein equations.

Let us compare this formulation with the Cartan-Weyl form of the gravity action for the usual tetrad (4 $\times$ 4 matrix). The action is
\begin{eqnarray}\label{S-Cartan-Weyl}                                      
& & S = \int e^\lambda_i e^\mu_k R_{\lambda \mu}{}^{ik} \sqrt{g} \d^4 x, ~~~ R_{\lambda \mu}{}^{ik} = \partial_\lambda \omega^{ik}_\mu - \partial_\mu \omega^{ik}_\lambda + ( \omega_\lambda \omega_\mu - \omega_\mu \omega_\lambda )^{ik}, \nonumber\\ & & g_{\lambda \mu} = e^i_\lambda e_{\mu i}, ~~~ i, k, l, \dots =1, \dots, 4.
\end{eqnarray}

\noindent The tetrad $e^\lambda_i$ can be considered as a particular case of the above $f_A^\lambda$ such that $f_A^\lambda = 0$ at $A > 4$. The equations of motion for the connection $\omega^{ik}_\lambda$ lead to
\begin{equation}\label{omega*ee}                                           
\omega_{\lambda ik} e^i_\mu e^k_\nu = \frac{1}{2} (T_{\mu, [\lambda \nu]} + T_{\nu, [\mu \lambda]} - T_{\lambda, [\nu \mu]}).
\end{equation}

\noindent Here again
\begin{equation}                                                           
T_{\lambda, [\mu \nu]} = e^i_\lambda (e_{\mu i, \nu} - e_{\nu i, \mu}).
\end{equation}

\noindent But now $T_{\lambda, [\mu \nu]} = 0$ means zero curvature $R_{\lambda \mu}{}^{ik}$, therefore nontrivial case requires $T_{\lambda, [\mu \nu]} \neq 0$. So we can compare both formulations in the following table.

\begin{flushleft}
\begin{tabular}{|l|c|c|}          \hline
~~~~~~~~~~formulation & Cartan-Weyl      & Faddeev            \\
variable  &      &  \\ \hline
tetrad &  $f^A_\lambda = 0 \mbox{~~at~} A > 4$   & $f^A_\lambda \neq 0 \mbox{~~at~} A > 4$ \\
          & (originally)     &                            \\ \hline
torsion  & $T_{\lambda, [\mu \nu ]} \neq 0$ & $T_{\lambda, [\mu \nu ]} = 0$ \\
          &      & (on eqs. of motion)               \\ \hline
          \end{tabular}
\unitlength 1pt
\begin{picture}(0,0)
\thicklines
\put(-323,68){\line(5,-2){107}}
\end{picture}
\end{flushleft}

\noindent It is seen that both formulations are in a sense mutually dual: a variable that is zero in one formulation is nonzero in another one, and vice versa. This is emphasized by the fact that the number of independent variables is the same in both cases: 4 $\times$ 10 = 40 components of $f^A_\lambda$ on one hand and 4 $\times$ 4 = 16 components of $e^i_\lambda$ plus 4 $\times$ 6 (the number of antisymmetric pairs $ik$) = 24 components of $\omega^{ik}_\lambda$, or again 40 on another hand. Therefore it is interesting to ask whether the formulation exists which in a sense contains these both. The formulation of interest should generalize these both and result in that one of interest under appropriate additional conditions. Faddeev formulation is already quite general one for it corresponds to $f^A_\lambda \neq 0$ at $A > 4$ and $T_{\lambda, [\mu \nu ]} \neq 0$ from the very beginning. As for the Cartan-Weyl formulation, it still admits generalization to such set of variables by simply generalizing the tetrad. To this end we rewrite the action (\ref{S-Cartan-Weyl}) by extending the region of values of the local vector index,
\begin{equation}\label{SSO(10)}                                            
S = \int f^\lambda_A f^\mu_B [\partial_\lambda \omega^{AB}_\mu - \partial_\mu \omega^{AB}_\lambda + ( \omega_\lambda \omega_\mu - \omega_\mu \omega_\lambda )^{AB}] \sqrt{g} \d^4 x.
\end{equation}

\noindent Here $\omega^{AB}_\lambda$ is SO(10) infinitesimal connection.

\section{Cartan-Weyl action for SO(10)}

We can perform the gauge transformation of the local frames which forces $f^\lambda_A$ have nonzero components only at $A = 1, 2, 3, 4$, i. e. have the sense of the usual tetrad. The equations of motion (\ref{ff*omega=dff}) below can be satisfied by $\omega_{\lambda AB}$ having nonzero components only at $A, B = 1, 2, 3, 4$. Thus, the action (\ref{SSO(10)}) can be reduced to the Hilbert-Einstein one just as the Cartan-Weyl action (\ref{S-Cartan-Weyl}). For what follows, however, it is of interest to get the same result without partial gauge fixing for $f^\lambda_A$.

The action (\ref{SSO(10)}) depends on $\omega_{\lambda AB}$ only through horizontal projections of $\omega_{\lambda AB}$ over one of SO(10) indices, $\omega_{\lambda BA} f_\mu^B$. This is evident for the bilinear over $\omega$ terms in (\ref{SSO(10)}). As for the terms with derivatives, we write, e. g.,
\begin{equation}\label{ffdomega}                                           
f^\lambda_A f^\mu_B \partial_\lambda \omega^{AB}_\mu = \partial_\lambda (f^\lambda_A f^\mu_B \omega^{AB}_\mu) - \omega^{AB}_\mu (\partial_\lambda f^\lambda_A) f^\mu_B - \omega^{AB}_\mu f^\lambda_A \partial_\lambda f^\mu_B.
\end{equation}

\noindent Here RHS contains $\omega_{\lambda AB}$ just in the form $\omega_{\lambda BA} f_\mu^B$.

In turn, decompose this horizontal projection into the horizontal and vertical com\-po\-nents over the remaining SO(10) index,
\begin{equation}\label{wf=hf+v}                                            
\omega_{\lambda BA} f^B_{\mu} = h_{\lambda \mu \nu} f^\nu_A + v_{\lambda \mu A}, ~~~ \Pi^{AB} v_{\lambda \mu B} = v^A_{\lambda \mu}.
\end{equation}

\noindent Evidently, $h_{\lambda \mu \nu} = - h_{\lambda \nu \mu}$.

Varying the action (\ref{SSO(10)}) w. r. t. $\omega^{AB}_{\lambda}$, we get the eqs. of motion for $\omega$,
\begin{equation}\label{ff*omega=dff}                                       
(f^\lambda_A f^{\mu C} - f^{\lambda C} f^\mu_A) \omega_{\mu CB} - (A \leftrightarrow B) = \frac{1}{\sqrt{g}} \partial_\mu [(f^\lambda_A f^\mu_B - f^\lambda_B f^\mu_A) \sqrt{g}].
\end{equation}

\noindent First, let us take vertical components of both sides of this equation (w. r. t., say, index $B$) and substitute the expansion (\ref{wf=hf+v}) for $\omega_{\lambda AB}$ here. This gives for the vertical part of the connection
\begin{equation}\label{v lambda mu A}                                      
v_{\lambda \mu}^A = \Pi^{AB} f_{\mu B, \lambda} \equiv b^A_{\mu \lambda}.
\end{equation}

Substitute the $v^A_{\lambda \mu}$ found back to (\ref{wf=hf+v}) and to (\ref{ff*omega=dff}). This gives eqs. for $h_{\lambda \mu \nu}$,
\begin{eqnarray}                                                           
& & \hspace{-10mm} f^\lambda_A h_\mu{}^\mu{}_\nu f^\nu_B - f^\mu_A h_\mu{}^\lambda{}_\nu f^\nu_B - (A \leftrightarrow B) = f^\lambda_A \Pi_{\| BC} f^{\mu C}_{, \, \mu} - f^\mu_A \Pi_{\| BC} f^{\lambda C}_{, \, \mu} - (A \leftrightarrow B) \nonumber \\ & & + (f^\lambda_A f^\mu_B - f^\lambda_B f^\mu_A) \partial_\mu \ln \sqrt{g}.
\end{eqnarray}

\noindent Here $\Pi_{\| AB} = \delta_{AB} - \Pi_{AB} = f^\lambda_A f_{\lambda B}$ is horizontal projector. Taking into account that
\begin{equation}                                                           
\partial_\mu \ln \sqrt{g} = f^\nu_C f^C_{\nu, \mu}
\end{equation}

\noindent and lowering $\lambda$, we get
\begin{equation}                                                           
f^\mu_A f^\nu_B (h_{\mu\nu\lambda} + h_{\nu\lambda\mu} + g_{\lambda \mu} h_\nu - g_{\nu \lambda} h_\mu ) = f^\mu_A f^\nu_B (T_{\lambda, [\nu \mu]} + g_{\lambda \mu} T_\nu - g_{\nu \lambda} T_\mu).
\end{equation}

\noindent Here $h_\lambda \equiv g^{\mu \nu} h_{\mu \nu \lambda}$, $T_\lambda \equiv g^{\mu \nu} T_{\mu, [\nu \lambda]}$. The system is equivalent to a smaller number of independent components,
\begin{equation}                                                           
h_{\mu\nu\lambda} + h_{\nu\lambda\mu} + g_{\lambda \mu} h_\nu - g_{\nu \lambda} h_\mu = T_{\lambda, [\nu \mu]} + g_{\lambda \mu} T_\nu - g_{\nu \lambda} T_\mu,
\end{equation}

\noindent and has formally the same solution for $h_{\lambda \mu \nu}$ as Cartan-Weyl formalism for $\omega_{\lambda ik} e^i_\mu e^k_\nu$ (\ref{omega*ee}),
\begin{equation}                                                           
h_{\lambda \mu \nu} = \frac{1}{2} (T_{\mu, [\lambda \nu]} + T_{\nu, [\mu \lambda]} - T_{\lambda, [\nu \mu]}).
\end{equation}

The curvature tensor for the connection $\omega_{\lambda AB}$
\begin{equation}                                                           
R_{\lambda \mu AB} = \partial_\lambda \omega_{\mu AB} - \partial_\mu \omega_{\lambda AB} + (\omega_\lambda \omega_\mu - \omega_\mu \omega_\lambda)_{AB}
\end{equation}

\noindent being horizontally projected over both SO(10) indices, i. e. $R_{\lambda \mu AB} f_\nu^A f_\rho^B$ does not depend on $\omega_{\lambda AB}$ at $A > 4$, $B >4$. It depends only on $\omega_{\lambda BA} f^B_\mu$. This follows upon rewriting the terms with derivatives like (\ref{ffdomega}). This gives
\begin{equation}                                                           
f_{\nu A} f_{\rho B} \partial_\lambda \omega_{\mu}^{ AB} = \partial_\lambda h_{\mu \nu \rho} + h_{\mu \rho \sigma} \Omega^{\sigma}{}_{\nu \lambda} - h_{\mu \nu \sigma} \Omega^{\sigma}{}_{\rho \lambda} + S_{\nu \rho, \lambda \mu}.
\end{equation}

\noindent Here $S_{\nu \rho, \lambda \mu}$ (\ref{S lambda mu nu rho}) arises from the vertical part $v_{\lambda \mu A}$ of the connection $\omega_{\lambda BA} f^B_\mu$. Also we find for the bilinear in $\omega$ part of the curvature tensor that
\begin{equation}                                                           
(\omega_\lambda \omega_\mu - \omega_\mu \omega_\lambda)^{AB} f_{\nu A} f_{\rho B} = - h_{\lambda \nu \sigma} h_{\mu \rho}{}^\sigma + h_{\mu \nu \sigma} h_{\lambda \rho}{}^\sigma - S_{\nu \rho, \lambda \mu}.
\end{equation}

\noindent Again, the last term here is due to $v_{\lambda \mu A}$. In overall,
\begin{eqnarray}\label{R=h+S}                                              
& & R_{\lambda \mu}^{ AB} f_{\nu A} f_{\rho B} = \partial_\lambda h_{\mu \nu \rho} - \partial_\mu h_{\lambda \nu \rho} + h_{\mu \nu \sigma} h_{\lambda \rho}{}^\sigma - h_{\lambda \nu \sigma} h_{\mu \rho}{}^\sigma \nonumber \\ & &  \hspace{10mm} + h_{\mu \rho \sigma} \Omega^{\sigma}{}_{\nu \lambda} - h_{\lambda \rho \sigma} \Omega^{\sigma}{}_{\nu \mu} + h_{\lambda \nu \sigma} \Omega^{\sigma}{}_{\rho \mu} - h_{\mu \nu \sigma} \Omega^{\sigma}{}_{\rho \lambda} \nonumber \\ & & \hspace{10mm} + S_{\nu \rho, \lambda \mu},
\end{eqnarray}

\noindent where the last term is due to $v_{\lambda \mu A}$. Using the identity \begin{equation}                                                           
h_{\lambda \mu \nu} = \Omega_{\nu, \mu \lambda} - \Gamma_{\nu, \mu \lambda}
\end{equation}

\noindent in eq. (\ref{R=h+S}) and taking into account that
\begin{equation}                                                           
\partial_\lambda \Omega_{\rho, \nu \mu} - \partial_\mu \Omega_{\rho, \nu \lambda} + \Omega_{\sigma, \rho \mu} \Omega^{\sigma}{}_{\nu \lambda} - \Omega_{\sigma, \rho \lambda} \Omega^{\sigma}{}_{\nu \mu} = - S_{\nu \rho, \lambda \mu}
\end{equation}

\noindent we see that the last term is canceled, and we are left with the Riemannian curvature tensor
\begin{equation}                                                           
R_{\lambda \mu AB} f_{\nu A} f_{\rho B} = g_{\rho \sigma} (\Gamma^\sigma_{\nu \lambda, \mu} - \Gamma^\sigma_{\nu \mu, \lambda} + \Gamma^\sigma_{\tau \mu} \Gamma^\tau_{\nu \lambda} - \Gamma^\sigma_{\tau \lambda} \Gamma^\tau_{\nu \mu}) = g_{\rho \sigma} R^{\sigma}_{\nu \mu \lambda} = R_{\nu \rho \lambda \mu} = R_{\lambda \mu \nu \rho}
\end{equation}

\noindent and, in particular, with the Hilbert-Einstein action
\begin{equation}\label{int R sqrt g d4x}                                   
\int R^{AB}_{\lambda \mu} f^\lambda_A f^\mu_B \sqrt{g} \d^4 x = \int R \sqrt{g} \d^4 x.
\end{equation}

\section{Representation for the Faddeev action}

Important point in the above derivation is occurrence of the Faddeev action for gravity (\ref{Fad action}) as contribution to (\ref{int R sqrt g d4x}) from the vertical components of $\omega_{\lambda BA} f^B_\mu$. The contributions of the horizontal and vertical components of $\omega_{\lambda BA} f^B_\mu$ to (\ref{SSO(10)}) do not mix. Therefore eqs. of motion for $\omega_{\lambda AB}$ can be written separately for the horizontal and vertical components. So to get purely Faddeev action we could impose conditions requiring vanishing the horizontal components of $\omega_{\lambda BA} f^B_\mu$. Taking these into account with the help of the Lagrange multipliers we can write out the representation for the Faddeev action a la Cartan-Weyl as
\begin{equation}\label{Faddeev-Cartan-Weyl}                                
S = \int [ R_{\lambda \mu}^{AB} (\omega ) + \Lambda^\nu_{[\lambda \mu]} \omega_{\nu}^{AB} ] f^\lambda_A f^\mu_B \sqrt{g} \d^4 x.
\end{equation}

\noindent Here $\Lambda^\nu_{[\lambda \mu]}$ are the Lagrange multipliers. This action possesses the global SO(10) sym\-met\-ry but not local one (due to the $\Lambda$-term). Note that redefining variables via scaling $f^\lambda_A g^{1/4} \equiv \tilde{f}^\lambda_A$ makes the action polynomial one.

\section{Barbero-Immirzi parameter}

Some natural generalization of the considered representation can be made. It is ana\-lo\-go\-us to generalization of the Cartan-Weyl form of the Hilbert-Einstein action (\ref{S-Cartan-Weyl}) via adding a term to it which vanishes on the eqs. of motion for connections \cite{Holst,Fat},
\begin{equation}                                                           
\int e^\lambda_i e^\mu_k R_{\lambda \mu}{}^{ik} (\omega ) \sqrt{g} \d^4 x \Longrightarrow \int \left ( e^\lambda_i e^\mu_k + \frac{1}{2 \gamma \sqrt{g}} \epsilon^{\lambda \mu \nu \rho} e_{\nu i} e_{\rho k} \right ) R_{\lambda \mu}{}^{ik} (\omega ) \sqrt{g} \d^4 x.
\end{equation}

\noindent Here $\gamma$ is a constant known as Barbero-Immirzi parameter \cite{Barb,Imm}. This can be imme\-di\-a\-te\-ly transferred to the considered SO(10) case (\ref{Faddeev-Cartan-Weyl}),
\begin{equation}                                                           
\label{S Imm full}
S = \int \left ( f^\lambda_A f^\mu_B + \frac{1}{2 \gamma \sqrt{g}} \epsilon^{\lambda \mu \nu \rho} f_{\nu A} f_{\rho B} \right ) R_{\lambda \mu}^{AB} (\omega ) \sqrt{g} \d^4 x + \int \Lambda^\nu_{[\lambda \mu]} \omega_{\nu}^{AB} f^\lambda_A f^\mu_B \sqrt{g} \d^4 x.
\end{equation}

Let us exclude $\omega_{\lambda AB}$. Applying variation $\Pi^{DB} \delta / \delta \omega^{AB}_\lambda$ to action we get the eqs. of motion
\begin{eqnarray}                                                           
& & \Pi^{DB} \left ( f^\lambda_A f^{\mu C} - f^{\lambda C} f^\mu_A + \frac{1}{\gamma \sqrt{g}} \epsilon^{\lambda \mu \nu \rho} f_{\nu A} f^C_\rho \right ) \omega_{\mu CB}
\phantom{- f^\lambda_B f^\mu_A) \sqrt{g} + \frac{1}{\gamma} \epsilon^{\lambda \mu \nu \rho} f_{\nu A} f_{\rho B}}
\nonumber \\ & &
\phantom{\Pi^{DB} ( f^\lambda_A f^{\mu C} - f^{\lambda C} f^\mu_A }
= \frac{1}{\sqrt{g}} \Pi^{DB} \partial_\mu \left [(f^\lambda_A f^\mu_B - f^\lambda_B f^\mu_A) \sqrt{g} + \frac{1}{\gamma} \epsilon^{\lambda \mu \nu \rho} f_{\nu A} f_{\rho B} \right ].
\end{eqnarray}

\noindent Solution for $\omega_{\lambda BA} f^B_\mu$ is the same as in the above case $1/ \gamma = 0$ and is $v_{\lambda \mu A}$ (\ref{v lambda mu A}) found above,
\begin{equation}                                                           
\omega_{\lambda BA} f^B_{\mu} = \Pi_{AB} f^B_{\mu, \lambda}.
\end{equation}

\noindent Substituting this back to (\ref{S Imm full}) we find the action
\begin{eqnarray}\label{Fad action+Imm term}                             
\int {\cal L} \d^4 x & = & \int \Pi^{AB} \left [ (f^\lambda_{A, \lambda} f^\mu_{B, \mu} - f^\lambda_{A, \mu} f^\mu_{B, \lambda}) \sqrt {g} - \frac{1}{\gamma} \epsilon^{\lambda \mu \nu \rho} f_{\lambda A, \mu} f_{\mu B, \rho} \right ] \d^4 x \\ \label{Fad action+Imm term 2} & = & \int g^{\lambda \nu} g^{\mu \rho} S_{\lambda \mu, \nu \rho} \sqrt {g} \d^4 x + \frac{1}{2 \gamma} \int \epsilon^{\lambda \mu \nu \rho} S_{\lambda \mu, \nu \rho} \d^4 x
\end{eqnarray}

\noindent which modifies (\ref{Fad action}) by adding a parity odd term.

Finally, write out the vertical components of the eqs. of motion for (\ref{Fad action+Imm term}). Applying variation $\Pi_{AB} \delta / \delta f^\lambda_B$ to action we get
\begin{equation}                                                           
b^\mu{}_{\mu A} T^\nu_{[\lambda \nu]} + b^\mu{}_{\lambda A} T^\nu_{[\nu \mu]} + b^\mu{}_{\nu A} T^\nu_{[\mu \lambda]} + \frac{\epsilon^{\tau \mu \nu \rho}}{2 \gamma \sqrt{g}} (g_{\lambda \sigma} g_{\kappa \tau} - g_{\lambda \tau} g_{\kappa \sigma}) b^{\kappa}{}_{\rho A} T^\sigma_{[\mu \nu]} = 0.
\end{equation}

\noindent This is modification of the above $V_{\lambda A} = 0$ with $V_{\lambda A}$ given by (\ref{V lambda A}). This modification seems to be not qualitatively crucial, and these equations still give $T^\lambda_{[\mu\nu]} = 0$ almost everywhere in the infinite-dimensional configuration superspace. Then again the cur\-va\-tu\-re tensor $S^\lambda_{\mu, \nu\rho}$ is Riemannian one $R^\lambda_{\mu, \nu\rho}$. The second term in (\ref{Fad action+Imm term 2}) is identically zero by properties of the Riemannian tensor, and we are left with purely Einstein action.

\section{Conclusion}

Thus we have studied the first order representation of the Faddeev gravity action in terms of the connections as additional variables. In this representation the action is polynomial. After excluding connection variables via equations of motion, we get genuine Faddeev action. The considered representation is analogous to the Cartan-Weyl form of the Einstein gravity action, now with gauge group SO(10), but is invariant only under the global SO(10), not under local one. Also we have studied the analog of the Barbero-Immirzi parameter considered in the literature in the Cartan-Weyl form of the Einstein gravity action. The analog of this parameter for our case defines a coefficient at certain parity-odd expression added to action in our representation. After excluding connection variables via equations of motion we get Faddeev action plus some parity-odd contribution. In turn, upon partial use of the equations of motion which reduces Faddeev action to the Hilbert-Einstein one, the parity-odd contribution vanishes, as it does in the case of genuine Barbero-Immirzi parameter.

The author thanks I.A. Taimanov who had attracted author's attention to the new formulation of gravity and Ya.V. Bazaikin for valuable discussion on this subject. The author is grateful to I.B. Khriplovich who has provided moral support, A.A. Pomeransky and A.S.Rudenko for discussion at a seminar, stimulating the writing of this article. The present work was supported in part by the Russian Foundation for Basic Research through Grants No. 09-01-00142-a, 11-02-00792-a and Grant \\ 14.740.11.0082 of federal program "personnel of innovational Russia".


\end{document}